\documentclass[prb,aps,superscriptaddress,amsmath,amssymb,showpacs,twocolumn,floatfix]{revtex4}

\usepackage{hyperref}
\usepackage{graphicx}
\newcommand{\rem}[1]{}

\begin{document}

\title{Mechanical properties of carbynes investigated by {\it ab initio}
  total-energy calculations}
\author{Ivano E. Castelli}
\affiliation{Center for Atomic-scale Materials Design,
Department of Physics, Technical University of Denmark,
DK-2800 Kongens Lyngby, Denmark}

\author{Paolo Salvestrini}
\affiliation{CNR-IFN - Istituto di Fotonica e Nanotecnologie, 
Sezione di Milano, Piazza L. Da Vinci 32, 20133 Milano, Italy}

\author{Nicola Manini}
\affiliation{ETSF and Dipartimento di Fisica, Universit\`a degli Studi di
  Milano, Via Celoria 16, 20133 Milano, Italy}

\begin{abstract}
As $sp$ carbon chains (carbynes) are relatively rigid molecular objects,
can we exploit them as construction elements in nanomechanics?
To answer this question, we investigate their remarkable mechanical
properties by {\it ab-initio} total-energy simulations.
In particular, we evaluate their linear response to small longitudinal and
bending deformations and their failure limits for longitudinal compression
and elongation.
\end{abstract}


\maketitle
\section{Introduction}

The rich chemistry of carbon is due to the capability of its electronic
configuration to adjust to different bonding situation.
Carbon atoms realize three main hybridization schemes of the valence
orbitals: $sp^3$, $sp^2$, and $sp$ hybridization.
$sp^3$ bonding tends to form three-dimensional networks, as in diamond and
the related amorphous structures.
$sp^2$-hybridized orbitals lead to two-dimensional networks, at the basis
of graphene, graphite, fullerenes, nanotubes, ribbon structures, and other
forms of wide current interest for their unique mechanical and electronic
properties.
$sp$-hybridized carbon forms linear structures (carbynes), which, compared
to their chemically more stable $sp^3$ (alkanes) and $sp^2$ (alkenes)
hydrogenated counterparts, tend to be rigid, thus potentially appealing as
backbones for molecular nanotechnology.

Carbynes of increasing length and varied terminations are being synthesized
\cite{Mohr03,Zhao03,Liu03,Inoue10} and characterized.\cite{Rice10}
At the same time, the fabrication of carbon atomic chains from stretched
nanotubes or graphene was achieved by controlled electron irradiation in
transmission electron microscopes.\cite{Troiani03,Jin09,Chuvilin09}
The remarkable robustness of these carbon chains under irradiation combined
with the ease of electron-beam fabrication at the nm scale can provide a
route to the synthesis of actual nano-devices based on carbynes.
%
Recent investigations have elucidated the prominent role of molecular
oxygen as a primary source of chemical degradation of $sp$ carbon
\cite{Moras11trib,Moras11}.
In a pure-carbon environment carbynes are far more stable
\cite{Casari04,Ravagnan07,Ravagnan09,Erdogan11,Castelli12}.

Effects of longitudinal straining
\cite{Hobi10,Cahangirov10,Topsakal10,Akdim11} and lateral bending
\cite{Hu11} were investigated very recently by {\it ab-initio} methods.
The outcome of those studies is that carbyne chains are at one time
extremely stiff against longitudinal straining and very soft against a
bending deformation, to the extent that even extreme bending affects only
moderately their bonding properties.\cite{Hu11}
In the present paper we focus on the mechanical properties of carbynes in
the perspective of exploiting them as construction materials for
nano-engineering.
We use {\it ab-initio} simulations \cite{Ravagnan09,Ataca11} to predict the
stiffness of the carbynes for both longitudinal strain and for bending, and
their ultimate tensile strength.
The main results of these simulations is that a sp carbon chain (and it
matters little whether it is of cumulenic -- all double bonds, or polyynic
-- with alternating single-triple bonds) exhibits not only an exceptional
mechanical stiffness against longitudinal deformations, but also a small
but nonzero rigidity against bending.
Accordingly, a C$_n$ chain resembles a thin beam characterized by a
finite rigidity against buckling, rather than a string.
We find that, e.g., a single-carbyne C$_8$ ``pillar'' can withstand a
modest but non-negligible axial compression force $\sim 0.7$~nN before
buckling.

\section{The model}\label{model:sec}

\begin{figure}
  \centerline{
    \includegraphics[width=80mm,angle=0,clip=]{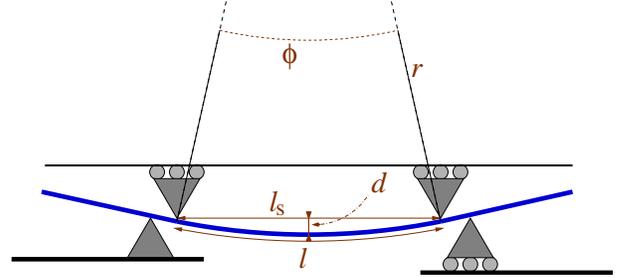}
  }
\caption{\label{model:fig}
 (Color online) A beam model used to describe the mechanical
  properties of a $sp$-carbon chain subject to straining.
 The scheme represents the relaxed configuration of the beam subjected to
 four lateral constraints, which induce two bending moments on the beam
 central section, which, as a result, bends as the arc of a circle.
 We identify the end-to-end distance $l_s$ of the deformed region; the
 curvilinear beam length $l$; the maximum deflection $d$ from straight
 geometry; the radius of curvature $r$; and the deflection angle $\phi$.
}
\end{figure}

In simulations, one can tune the carbyne length by fixing the position of
the end carbon atoms of a selected chain section, and leave all other atoms
free to relax to their equilibrium position, compatible with the imposed
strain.
As well know to civil engineers, as soon as the compressive stress exceeds
a critical value, the straight configuration of a beam subject to a purely
axial strain $\varepsilon <0$ turns unstable and ``buckles'' spontaneously
to a curved arc.
In the simulations of highly strained configurations, it is straightforward
to address either the metastable straight configuration, or the stable
buckled configuration investigated in Ref.~\onlinecite{Hu11}.

To characterize the mechanical properties of a carbyne segment, and in
particular to evaluate the buckling transition, one needs to identify the
correct linear-response parameteres in a simple model where the effects of
different kinds of deformations can be decoupled and studied separately.
The simplest continuum model describes a carbyne as a strained thin beam of
a homogeneous material, thereby ignoring its discrete atomic structure.
Three basic deformations can be applied to the beam: longitudinal traction
or compression; lateral bending, as sketched in Fig.~\ref{model:fig};
torsion around its axis.
Correspondingly, the elastic deformation energy can be decomposed as:
\begin{equation}\label{Eel}
E_{\rm el} = E_{\rm tens}  + E_{\rm bend} + E_{\rm tors}
\,,
\end{equation} 
where $E_{\rm tens}$ is the tensile energy increase due to longitudinal
elongation/shortening, $E_{\rm bend}$ is the energy due to the pure flexion
of the beam, and $E_{\rm tors}$ is the energy increase due to the twisting
of the beam.
Remarkably, even this third term is important for the essentially 1D
cumulenes.\cite{Ravagnan09}
In the present work we will however mostly ignore torsion, and concentrate
on the first two terms.

In the linear-response (small-strain) regime, we write the tensile energy
in terms of the strained length $l$ relative to the fully relaxed length
$l_0$ as
\begin{equation}\label{Etens}
E_{\rm tens}  = \frac{1}{2} \frac {\chi}{l_0} \left( l - l_0 \right)^2
,
\end{equation} 
where $\chi/l_0$ is the elastic stiffness of the beam, and $\chi$ is the
length-independent beam elastic tension: in the continuum mechanics of 3D
solids $\chi ={\cal E}\cdot A$, the product of the material's elasticity
modulus times the beam transverse area.

Figure~\ref{model:fig} indicates the relevant geometrical quantities
characterizing the pure-bending deformations of a carbyne of fully relaxed
length $l_0$.
In a circular-arc geometry ensuing from the application of a small bending
moment $M$ at the beam ends, the bending energy can be derived by the
linear relation of the deflection angle $\phi$ with the applied bending
moment:
\begin{equation}\label{anglemoment}
\phi =
\frac{M l}{g}
\,,
\end{equation} 
Here $g$ is the length-independent bending stiffness of the beam: in 3D
continuum mechanichs $g=E\cdot I$, the product of the elasticity modulus
times the second moment of area of the beam cross section. (The dimension
of $g$ is force$\times$area.)
For a circular deformation, the bending energy $E_{\rm bend}$ is therefore
\begin{equation}\label{Ebend}
E_{\rm bend} = \frac{1}{2} M \phi = g \frac{\phi^2}{2 l}
\,.
\end{equation}

It is convenient to express all quantities in terms of clearly defined
geometrical quantities, e.g.\ the distance $l_s$ between the end points of
the chain and the the maximum deflection $d$ from straight geometry.
We have:
\begin{eqnarray}\label{angle}
\phi &=&
4 \arctan \frac{2 d}{l_s}
\\\label{length}
l &=& \phi \, r = \frac{\phi}{2 \sin \frac{\phi}{2}}\,l_s
\,.
\end{eqnarray} 
With the implicit substitution of the expressions (\ref{angle}) and
(\ref{length}), the elastic energy of the beam stretched and/or bent in a
circular arc is therefore
\begin{equation}\label{Emech}
  E_{\rm mech}=E_{\rm tens} +E_{\rm bend} =
  \frac{1}{2} \frac{\chi}{l_0}
  \left( l  - l_0 \right)^2 + g \frac{\phi^2}{2 l}
\,.
\end{equation}

\section{Calculations}\label{results:sec}

We determine the linear-response parameters $\chi$ and $g$ by means of
simulations based on the density functional theory (DFT) in the local
density approximation (LDA).
The time-honored LDA is one in many functionals being used for current DFT
studies of molecular systems: other functionals often improve one or
another of the systematic defects of LDA (underestimation of the energy
gap, small overbinding and overestimation of the vibrational frequencies),
but to date no functional is universally accepted to provide systematically
better accuracy than LDA for all properties of arbitrary systems.
For a covalent system of $s$ and $p$ electrons as the one studied here, LDA
is appropriate, and we expect our results to change by a few percent at
most if the calculations were repeated using some other popular
functional.\cite{B3LYP,PBE96,Xu04}

We compute the total adiabatic energies by means of the plane-waves DFT
code Quantum Espresso.\cite{espresso2009}\,%
\footnote{
To compute reliable small energy differences, we impose severe
self-consistency requirements: we accept the electronic-structure self
consistency when the total adiabatic energy is converged to better than
$10^{-10}$~Ry$\simeq 1$~peV, and push atomic coordinates relaxation until
all force components are smaller than $10^{-5}$~Ry$/a_0 \simeq 0.4$~pN.
We use ultrasoft pseudopotentials,\cite{Vanderbilt90,Favot99} for which a
moderate cutoff for the wave function/charge density of $30/240$~Ry is
sufficient.
To address isolated molecules in the repeated-cell geometry implied by
plane waves, we make sure that at least $1$~nm of vacuum separates all
atoms in adjacent periodic images.
}
We consider the two different limiting structures of carbynes: polyynes (or
$\alpha-$carbynes) with alternating single--triple bonds
(\dots$-$C$\equiv$C$-$C$\equiv$C$-$\dots), contrasted to cumulenes or
$\beta-$carbynes, with nearly-equal-length (double) bonds
(\dots=C=C=\dots).
A typical polyynic chain is obtained when each end carbon atom forms a
single bond with a ligand (e.g.\ a hydrogen atom, as in diacetylene), while
a typical cumulene is obtained when each end carbon forms a double bond,
for example to a CH$_2$ group, as in ethylene.

We focus on a C$_8$ chain segment, and compare its mechanical response in
its HC$_2$-C$_8$-C$_2$H polyynic and H$_2$C-C$_8$-CH$_2$ cumulenic
realizations.
For the polyyne, we select hydrogenacetylide terminations rather than a
single hydrogen because the HC$_2$- group allows us to impose a bending
moment at the ends of the C$_8$ chain as in Fig.~\ref{model:fig}, by
displacing a C atom along the chain continuation rather, than a chemically
different H atom, thereby probing the intrinsic properties of carbyne,
rather than those of a specifically-terminated compound.
We also consider C$_{n}$ polyyne of different lengths, to check for size
effects, and a $90^\circ$ end-twisted C$_8$ cumulene, which we must treat
within the local spin-density approximation (LSDA) since twisting induces a
peculiar total spin-1 electronic state, with a twofold-degenerate level
occupied by two parallel-spin electrons.\cite{Ravagnan09}
We do not attempt any serious size scaling, for two main reasons: (i) very
long carbynes are mechanically unstable (thus unsuitable for nanomechanical
applications) anyway, and (ii) the long-range interaction effects
demonstrated recently \cite{Cahangirov10} would make a proper {\it
  ab-initio} size-scaling determination of the mechanical (in particular
bending) properties of long carbynes prohibitively expensive.
%
In the present work we do not deal with the exotic odd-$n$ C$_n$ chains,
which are chemically less stable and more difficult to
synthesize.\cite{Chalifoux09,Cataldo10}

\begin{figure}
\centerline{
  \includegraphics[width=80mm,angle=0,clip=]{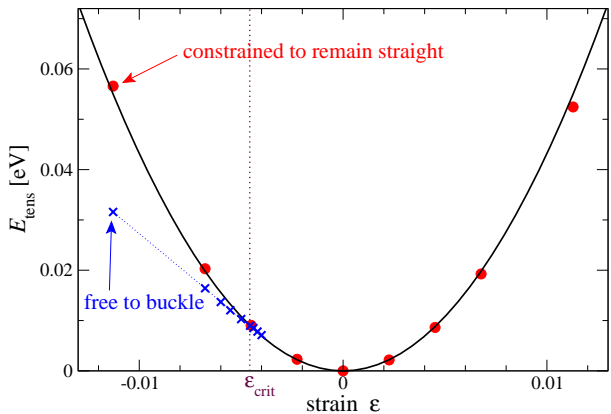}
}
\caption{\label{kfit:fig}
  (Color online) The DFT-LDA tensile excess energy computed for the
  strained straight HC$_2$-C$_8$-C$_2$H polyyne (bullets), as a
  function of $\varepsilon = (l-l_0)/l_0$.
  Solid line: a quadratic fit of Eq.~(\ref{Etens}), restricted to the 7
  data points within $\pm 0.8\%$ strain, where deviations from the
  linear-response regime are negligible.
  Crosses: the total excess energy without the constraint that the chain
  should remain straight: buckling starts to build up for strain exceeding
  $\varepsilon_{\rm crit}$.
  The very similar data and fit for the cumulene H$_2$C-C$_8$-CH$_2$ is not
  reported for clarity's sake.
}
\end{figure}

To evaluate $\chi$, starting from the fully relaxed carbyne we stretch or
compress a straight C$_8$ chain section by fixing the positions of the end
carbon atoms, and letting all other atoms free to relax to their resulting
equilibrium positions under stress.
We extract an estimate of $\chi$ by fitting Eq.~\eqref{Etens} to the
resulting values of the excess energy as a function of the chain
elongation, as reported in Fig.~\ref{kfit:fig}.

\begin{figure}
\centerline{
  \includegraphics[width=80mm,angle=0,clip=]{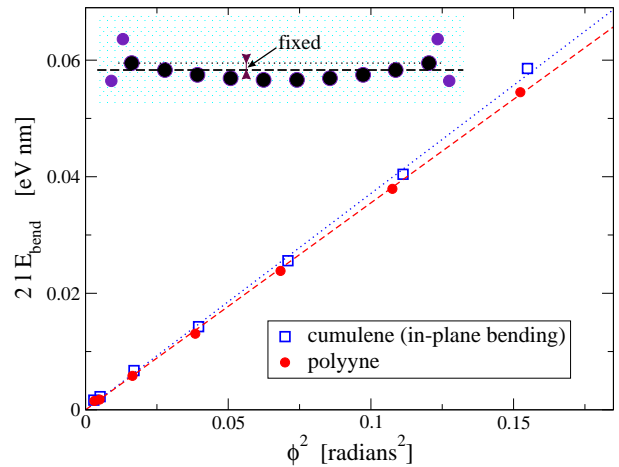}
}
\caption{\label{gfit:fig}
  (Color online) The bending energy $E_{\rm bend}=E_{\rm mech}-E_{\rm
    tens}$, see Eq.~\eqref{Emech}, as a function of the squared
  bending angle of a -C$_8$- chains.
  Here $E_{\rm mech}$ is the excess energy in the bent configuration
  induced by the 4-point fixed lateral displacement (at most 30~pm)
  sketched in Fig.~\ref{model:fig}, $E_{\rm tens}$ is the value obtained by
  Eq.~\eqref{Etens}, using $\chi$ as obtained in the straight-chain
  calculation of Fig.~\ref{kfit:fig}.
  Symbols are the DFT values, lines are linear fits, whose slopes,
  according to Eq.~\eqref{Ebend}, represent the values of $g$.
  Polyynic chain: circles and dashed line; in-plane bent cumulenic chain:
  squares and dotted line; the out-of-plane bent cumulenic chain results in
  very similar energies and is not shown for clarity.
  Inset: an example of fully relaxed in-plane bent configuration of the
  cumulene H$_2$C-C$_8$-CH$_2$.
}
\end{figure}

\begin{table*} 
\begin{center}
\begin {tabular}{l|c|c|c|c|c|c|c}
\hline \hline
 &HC$_2$-C$_4$-C$_2$H
  &HC$_2$-C$_8$-C$_2$H
   &\multicolumn{2}{c|}{H$_2$C-C$_8$-CH$_2$}
     & H$_2$C-C$_8$-CH$_2$
      &HC$_2$-C$_{12}$-C$_2$H
       &HC$_2$-C$_{16}$-C$_2$H
\\
 &(polyyne)
  &(polyyne)
   &\multicolumn{2}{c|}{(cumulene)}
     & (cumulene)
      & (polyyne)	
       & (polyyne)	
\\
 &
  &
   & \multicolumn{2}{c|}{(planar)}
     & ($90^\circ$-twisted)
      &
       &
\\
\hline
$l_0$ (C$_n$) [pm]
 &377
  &886
   &\multicolumn{2}{c|}{888}
     &889
      &1395
       &1903
\\
$\varepsilon_{\rm ult}$
 &16.2\%
  &16.3\%
   &\multicolumn{2}{c|}{17.9\%}
     &16.8\%
      &16.4\%
       &16.8\%
\\
$F_{\rm ult}$ [nN]
 &12.7
  &12.3
   &\multicolumn{2}{c|}{14.7}
     &12.7
      &12.1
       &12.1
\\
$\chi$ [nN]
 &150
  &156
   &\multicolumn{2}{c|}{151}
     &148
      &149
       &148
\\
 &
  &
   & out-of-plane
    & in-plane
     &
      &
       &
\\
$g$ [nN(nm)$^2$]
 &$7.91~10^{-2}$
  &$5.69~10^{-2}$
   &$5.68~10^{-2}$
    & $5.95~10^{-2}$
     &$5.62~10^{-2}$
      &$5.11~10^{-2}$
       &$5.03~10^{-2}$
\\
$\varepsilon_{\rm crit}$
 &$-3.7$\%
  &$-0.458$\%
   &$-0.470$\%
    & $-0.493$\%
     &$-0.473$\%
      &$-0.174$\%
       &$-0.093$\%
\\
$F_{\rm crit}$ [nN]
 &5.49
  &0.715
   &0.710
    &0.744
     &0.702
      &0.259
       &0.137
\\
\hline \hline
\end{tabular}
\end{center}
\caption{\label{values:tab}
The computed mechanical properties of selected carbynes.
The elastic tension $\chi$ and bending stiffness $g$ define the elastic
response, Eq.~\eqref{Emech}.
The ultimate strain $\varepsilon_{\rm ult}$ and tension $F_{\rm ult}$
define the maximum stretching that the beam can stand before it breaks.
For a C$_8$ chain section with both ends pinned (hinged, free to rotate),
$F_{\rm crit}$ measures the minimum compressive force (with a corresponding
strain $\varepsilon_{\rm crit}=(l_{s}^{\rm crit} -l_0)/{l_0}$ relative to
the reported equilibrium length $l_0$) that must be applied to induce
buckling.
}
\end{table*}

To evaluate the bending stiffness $g$, we force the chain to bend by
imposing a small lateral displacement of the two atoms adjacent to the
C$_8$ section of the chain, while keeping the first and last atom in the
chain section bounded to remain laterally undisplaced, following the scheme
of Figs.~\ref{model:fig} and \ref{gfit:fig} (inset).
We then let all other degrees of freedom (including the axial position of
the laterally bounded atoms) of the chain relax.
The C$_8$ section relaxes to form a circular arc, which we fit \footnote{
  We fit the value of $d$ by minimizing the squared distance of the relaxed
  atomic positions in the $x-z$ plane from the circular arc
  $x=d-r+(r^2-z^2)^{1/2}$, whose radius $r=(l_s^2+4 d^2)/(8 d)$ is
  determined by taking into account the separation $l_s$ of the terminal
  atoms of the C$_8$ chain section, namely those kept at $x=0$.
}
to extract the geometrical quantities indicated in Fig.~\ref{model:fig}.
Note that this procedure differs substantially from the fixed-end approach
of Ref.~\onlinecite{Hu11}, where longitudinal and bending strains are
applied at the same time, and deviations extend well outside the
linear-response region.
In the language of that work, the maximum arc-chord ratio in our
calculations is $l/l_s=1.0064$.
We repeat this procedure for several values of the fixed lateral
displacements.
With the computed DFT energies and structures, we extract the values of $g$
by fitting the angular dependency of $E_{\rm bend}$ in Eq.~\eqref{Ebend},
see Fig.~\ref{gfit:fig}.

\begin{figure}
\centerline{
  \includegraphics[width=80mm,angle=0,clip=]{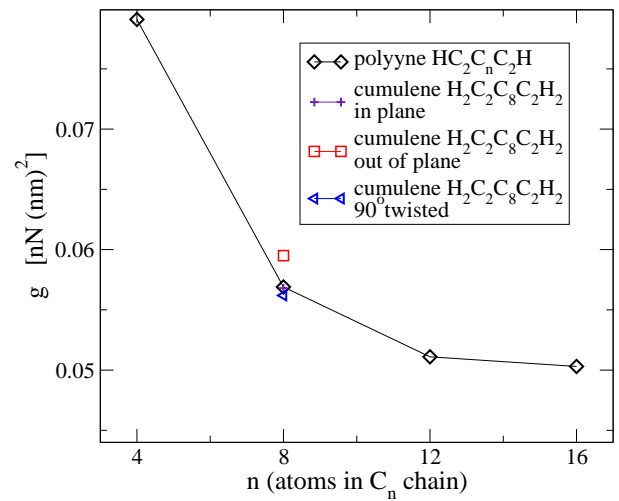}
}
\caption{\label{gscale:fig}
  (Color online) The chain-length dependency of the computed bending
  stiffness $g$, as reported in Table~\ref{values:tab}.
}
\end{figure}

The resulting elasticity parameters $\chi$ and $g$ are collected in
Table~\ref{values:tab}.
Despite the different bonding configuration of polyyne and cumulene, the
overall longitudinal elasticity $\chi$ values are remarkably similar.
The value of $\chi$ is compatible with a sound velocity
$v_s=(\chi/\mu)^{1/2}\simeq 31.5$~km/s for long-wavelength longitudinal
vibrations ($\mu$ is the linear mass density).
In view of the long-range interactions demonstrated recently
\cite{Cahangirov10}, the weak dependence of $\chi$ on the chain length is
remarkable.
The bending stiffness $g$ is comparatively more sensitive to the chain
length, see also Fig.~\ref{gscale:fig}.
If this velocity is used to evaluate the highest-frequency
Brillouin-zone--boundary phonon in a simplest mass-spring chain, one
obtains a frequency $\nu_{\rm BZb}=v_s/(\pi d_{\rm CC})\simeq 79$~THz,
i.e.\ a wave number $\sim 2600$~cm$^{-1}$, expectedly slightly above the
region of the observed high-frequency Raman and infrared modes emerging as
the characteristic spectroscopic signatures of the
carbynes.\cite{Ravagnan07,Ravagnan09,Onida10,Cataldo10,Innocenti10}

The cumulene is anisotropic: one can make it bend either within the
molecular plane, as sketched in the inset of Fig.~\ref{gfit:fig}, or
perpendicularly to it, resulting in differently oriented double bonds
\cite{Manini10} being perturbed.
We find that bending is significantly softer for the out-of-plane
distortion than for the in-plane one, with a $\sim 5$\% anisotropy.

The proposed cosine expression for the strain energy given in Eq.~(9) of
Ref.~\onlinecite{Hu11} is compatible with our Eq.~\eqref{Ebend} for small
curvature (although a small amount of longitudinal strain energy is
neglected there).
The relation $g \simeq d_{\rm CC} F_d$ obtained based on the average bond
length $d_{\rm CC}=129$~pm and orientation-dependent bonding strength
$F_d=2.54$~eV as obtained in Ref.~\onlinecite{Hu11} yields
$g=5.26\,10^{-2}$~nN/(nm)$^2$, in good agreement with our result, especially
with the C$_{12}$ chain value.
This agreement is even more surprising in view of the extreme deformations
introduced in Ref.~\onlinecite{Hu11}, and indicates that the linear-response
coefficient of Eq.~\eqref{Ebend} extends well outside its validity region,
provided that the cosine espression Eq.~(9) of that work is adopted.

According to elementary continuum mechanics,\cite{Timoshenko61} the
critical axial load that a beam can sustain in its straight configuration
before the onset of the buckling instability is given by Euler's formula
\begin{equation}\label{criticalforce}
F_{\rm crit} = \pi^2 \,\frac{g}{l^2}
\,,
\end{equation}
corresponding to a critical strain
\begin{equation}\label{criticalstrain}
\varepsilon_{\rm crit}
=\frac{l_{s}^{\rm crit} - l_0}{l_0}
\simeq
-\frac{F_{\rm crit}}{\chi}
Ã¸\simeq
-\pi^2\,\frac{g}{\chi l_0^2}
\,,
\end{equation}
for a critical sinusoidal lateral deformation profile.
By substituting the computed values for $\chi$ and $g$, we find that the
C$_8$ carbyne sustains a critical strain of nearly $-0.5$\%, under the action
of a critical force of approximately $0.7$~nN, before buckling, as reported
in Table~\ref{values:tab}.
The corresponding critical strain and force for cumulene (in the softer
out-of-plane direction) are essentially equal.
The computed values are relevant when both ends of C$_8$ are pinned
(hinged, free to rotate), while if both ends were frozen, the critical
force and strain would be four times larger.
Likewise, following Eqs.~\eqref{criticalforce} and \eqref{criticalstrain},
$F_{\rm crit}$ decays rapidly with size.
For example, for C$_{16}$, whose length is approximately 2.15 times that of
C$_8$, $F_{\rm crit} = 0.14$~nN, i.e.\ approximately 20\% of the C$_8$
value only.
Direct calculation done for the C$_{12}$ polyyne in HC$_2$-C$_{12}$-C$_2$H
agrees with this scaling, with a small deviation due mostly to a smaller
value of $g$, relative to the C$_8$ value (while $\chi$ is practically
coinciding, see Table~\ref{values:tab}).

\section{Discussion and Conclusion}

We find it rather surprising that, despite the presence of in principle
softer single bonds, the polyynic chain is essentially as hard to compress
as double-bonds--based cumulene.
The same essentially equivalent stiffness of polyyne and cumulene is found
against bending deformations, but here the result is less unexpected.
Even a maximally twisted cumulene in a high-spin electronic state exhibits
very similar mechanical properties, see Table~\ref{values:tab}.

We check the obtained buckling critical point for possible effects of chain
discreteness or anharmonicity, by running actual simulations.
We compare pairs of DFT simulations at fixed strain: one bound to the
linear configuration as in Fig.~\ref{kfit:fig}, and one starting off in a
slightly curved geometry.
In the curved simulations, below the buckling instability, the relaxing
chain goes back to straight, and recovers the same energy and outward force
that the chain produces on the pinned end atoms as in the straight
calculation.
In contrast, above the buckling instability, a curved shape stabilizes,
with a net decrease in total energy and in the force acting on the pinned
end atoms.
Crosses in Fig.~\ref{kfit:fig} report the fixed-end excess energy of the
polyynic chain when allowed to relax in a buckled geometry.
We find that the actual buckling instability occurs very close to the
linear-response continuum-model value $\varepsilon_{\rm crit}$ of
Eq.~\eqref{criticalstrain}.
Relaxations near instability are rather delicate, since equilibrium is
almost indifferent, providing a manifold of almost-equivalent geometrical
configurations which give a hard time to the optimization algorithm.

The present work demonstrates that it is possible to go a long way in
describing an immaterialy thin object such as a monoatomic carbon chain
with the mechanics of bulk construction elements, not unlike it was done
for 2D graphene in Refs.~\onlinecite{Cadelano09,Cadelano10}.
Eventually, carbyne chains turn out rigid enough that the 886~pm-long
free-standing C$_8$ polyyne can sustain a compressive strain of nearly
0.5\% before buckling.
On the tensile side, we extend our calculations well outside the
linear-response region, to estimate the ultimate tensile strength of
carbynes.
The ultimate tension $F_{\rm ult}$ is computed as the maximum force that
the carbyne section produces in sustaining an externally imposed
longitudinal strain $\varepsilon_{\rm ult}>0$.
The result in the 10~nN region agrees with previous determinations
\cite{Cahangirov10,Mazilova10} (but disagrees significantly with the
Tersoff-Brenner--model determination of Ref.~\onlinecite{Ragab11}), and
indicates that a single molecular chain could be used as a rope to lift a
mass as heavy as one microgram without breaking!
%
%
If one could pack carbynes with a lateral density of one per
$A=0.2\times0.2$~nm$^2$ cross-section, one would obtain a material
characterized by a remarkable ultimate tensile strength of the order
300~GPa, comparable to that of carbon nanotubes.\cite{Yu00}

Despite the remarkable mechanical properties of carbynes, their chemical
reactivity makes them unsuitable to many applications in real world
mechanical situations, where carbon fiber or carbon nanotubes provide
superior stability with comparable mechanical properties.
However in clean well-isolated nano-engineered devices one could envisage
that the usage of C$_n$ chains as structural or elastic elements may lead
to consistent advantages over traditional solutions.
For example, the soft bending degree of freedom could be exploited for the
construction of sensitive accelerometers.
Indeed, according to our evaluation of $g$, a 0.1~$\mu$g mass hanging at
the end of a 37~nm-long C$_{30}$ chain (with the other end bonded to a
fixed substrate) would deflect laterally by as much as 1~nm under an
acceleration of 0.01~m\,s$^{-2}$, i.e.\ one thousandth of the Earth
gravitational field.

\acknowledgments
We acknowledge useful discussion with G. Onida and L. Ravagnan.


\bibliographystyle{unsrt}

\end{document}